\documentclass[aps,twocolumn,groupedaddress,prb,floatfix,showpacs]{revtex4}
\usepackage{graphicx}
\usepackage{dcolumn}
\usepackage{bm}
\usepackage{amsmath}
\usepackage{verbatim}

\begin{document}

\title{Dielectric capping effects on binary and ternary topological insulator surface states}

\author{Jiwon Chang}
\email{jiwon.chang@utexas.edu}
\author{Priyamvada Jadaun}
\author{Leonard F. Register}
\author{Sanjay K. Banerjee}
\author{Bhagawan Sahu}
\affiliation{
Microelectronics Research Center, The University of Texas at Austin, Austin Texas 78758
}

\date{\today}

\begin{abstract}
Using a density functional based electronic structure method, we study the effect of crystalline dielectrics on the metallic surface states of Bismuth- and chalcogen-based binary and ternary three dimensional topological insulator (TI) thin films.  Crystalline quartz (SiO$_2$) and boron nitride (BN) dielectrics were considered. Crystalline approximation to the amorphous quartz allows to study the effect of oxygen coverage or environmental effects on the surface states degradation which has gained attention recently in the experimental community. We considered both {\it symmetric} and {\it asymmetric} dielectric cappings to the sufaces of TI thin films. Our studies suggest that BN and quartz cappings have negligible effects on the Dirac cone surface states of both binary and ternary TIs, except in the case of an oxygen-terminated quartz surface. Dangling bond states of oxygens in oxygen-terminated quartz dominate the region close to Fermi level, thereby distorting the TI Dirac cone feature and burying the Dirac point in the quartz valence band region. Passivating the oxygen-terminated surface with atomic hydrogen removes these dangling bond states from the Fermi surface region, and consequently the clear Dirac cone is recovered. Our results are consistent with recent experimental studies of TI surface degradation in the presence of oxygen coverage.         
\end{abstract}

\maketitle

\section{Introduction}

Three dimensional (3D) topological band insulator (TI) Bi$_2$X$_3$ (X=Se, Te) and its ternary counterparts have attracted considerable attention from the condensed matter physics community because of the relatively simple crystal structure that hosts novel surface states\cite{zhang1,zahid1}. and their unusual responses to external fields. Many more 3D TI materials have been predicted\cite{new3d1,new3d2,new3d3,new3d4,new3d5} by now, and quest for studying their novel surface state properties, in isolation as well as in presence of other materials, has increased in recent years. The 3D TI surface states are time-reversal symmetric (TRS) at high-symmetry points in the momentum space and are therefore protected against perturbations which cannot break TRS such as non-magnetic impurities or adatoms. Such novel properties have caused excitement in the electron device community as well because it can be a potential alternative to graphene as a channel material in field effect transistors. The advances in understanding of structural, electronic, magnetic and transport properties of 3D TI, made possible by both experimental and theoretical studies\cite{zhang2,Burkov,Hor}, can provide important informations for novel applications. Beside the fundamental studies of these model 3D TI materials, theoretically, no studies of dielectric effects on the surface states have been performed so far from first principles. We address dielectric effects on the TI surface states using {\it ab-initio} density functional theory (DFT) and semi-local density approximation\cite{perdew}. Two crystalline dielectrics were considered: quartz (SiO$_2$) and boron nitride (BN). Both binary TI Bi$_2$Se$_3$ and ternary TIs Bi$_2$Se$_2$Te and Bi$_2$Te$_2$Se were considered for this study. Our studies suggest that neither of the two dielectrics has an effect on the Dirac cone except the oxygen-terminated quartz. Under the environment of oxygen dangling bond, the Dirac cone on the TI surface is buried inside the quartz valence band continuum with the oxygen dangling bond states occupying the region around the Fermi level. Surface passivation by atomic hydrogen pushes the dangling bond states from the Fermi surface region down below Fermi level and a clear Dirac cone emerges at the Fermi level. These findings are consistent with recent experimental studies of surface degradation effects in the presence of oxygen\cite{shen}.

We begin by describing the thin film structures of both binary and ternary TIs, built from their bulk hexagonal structures, and the computational method in section II. In section III, we present the dielectric capping effects, by BN and SiO$_2$, on their metallic surface states. Both {\it symmetric} and {\it asymmetric} cappings are explored. Finally we present our summary and conclusions.

\section{Isolated Thin Films and Computational Approach}

\begin{figure}[ht!]
\scalebox{0.42}{\includegraphics{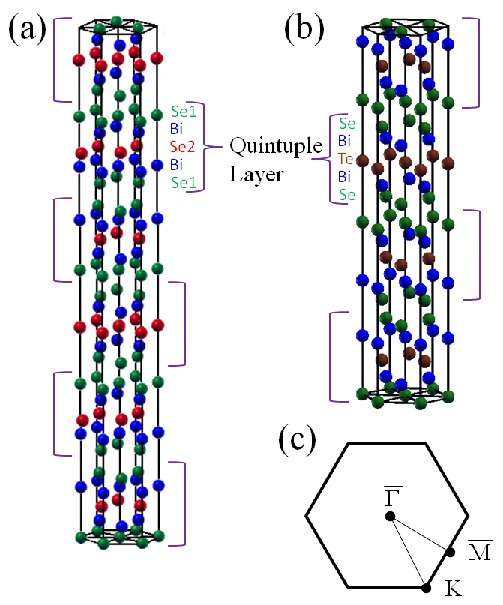}}
\caption{ (Color online) Schematic diagram of (a) 6QLs Bi$_2$Se$_3$ and (b) 4QLs Bi$_2$Se$_2$Te thin film structures obtained by stacking QLs along the {\it z}-direction. (c) Two-dimensional Brillouin zone (BZ) of the (111) surface of thin film TIs with three time-reversal invariant points $\bar{\Gamma}$, \={M}, and \={K}.}
\label{fig:Fig1}
\end{figure}

The thin film structures for both binary and ternary TIs are constructed by stacking up several quintuple layers (or QLs: 1QL=5 atomic layers) along the crystallographic {\it z}-direction with a vacuum region of 3 nm which forms the supercell in the DFT calculation. The first principle calculations were performed using the OPENMX code\cite{Openmx1}, based on a linear combination of pseudoatomic orbital (PAO) method\cite{Openmx2}. The pseudopotentials were generated from full relativistic calculations, and the generalized gradient approximation\cite{perdew} was applied for the exchange-correlation potential. The kinetic energy cut-off of 180 Rydberg and the {\bf k}-point mesh of 7 $\times$ 7 $\times$ 1 for Brillouin zone (BZ) integration (Fig. 1(c)) were used. The basis sets were carefully chosen to reproduce previous bulk and thin film calculations. The cut-off, {\bf k}-point mesh and vacuum region were optimized to guarantee the convergence of the results and their agreement with previous theoretical calculations. 

For the binary TI, previous theoretical studies of bulk Bi$_2$Se$_3$ found the computed lattice parameters close to the experimental values\cite{zhang2}, so we used the experimental lattice parameters {\it a}=0.41388 nm and {\it c}=2.8633 nm in the hexagonal unit cell and optimized bulk atomic positions to build the thin film structure. For the ternary TI Bi$_2$Se$_2$Te (Bi$_2$Te$_2$Se), we built thin film structures with the bulk hexagonal unit cell lattice parameters {\it a}=0.422 nm (0.428 nm) and {\it c}=2.92 nm (2.99 nm) in the previous theoretical calculation\cite{jiwon}, and optimized the thin film structure by letting atoms move along {\it z}-direction. We optimized thin film structures only for ternary TIs, since atomic relaxations significantly affect the band structure of thin film Bi$_2$Se$_2$Te around the Dirac point\cite{jiwon}. We considered the thin film of 6QLs and 4QLs for binary and ternary TIs, respectively, since previous studies on binary\cite{liu,zhang3} and ternary\cite{jiwon} TIs suggest that 6QLs and 4QLs are the minimum thicknesses to maintain the gapless surface state at the Dirac point. We also obtained the same critical thickness values by our own calculations. In the calculated band structure of 6QLs Bi$_2$Se$_3$, we could observe the Dirac cone within a bulk gap of 0.262 eV. For the ternary TIs Bi$_2$Se$_2$Te and Bi$_2$Te$_2$Se, Dirac cone surface states reside inside the bulk gap of 0.204 eV and 0.325 eV, respectively.

\begin{figure}[ht!]
\scalebox{0.53}{\includegraphics{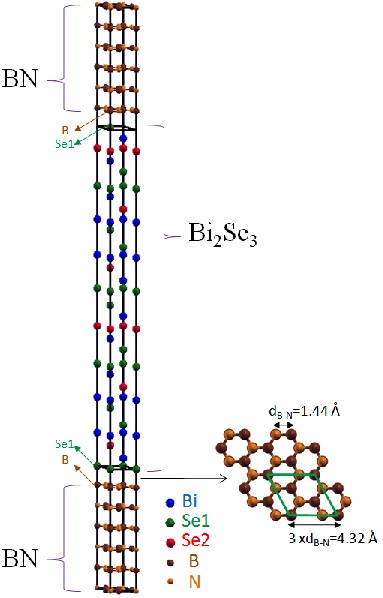}}
\caption{(Color online) Schematic diagram of the supercell structure of BN and 6QLs Bi$_2$Se$_3$. The colors corresponding to the atoms are labled on the bottom right of supercell. The green and brown arrows at the top and bottom interfaces indicate B on the top of Se. The top view of bottom interfacial atomic layers is shown on the bottom right of the supercell structure. The 1$\times$1 Bi$_2$Se$_3$ cell (green rhombus) is matched with 3d$_{B-N}$$\times$3d$_{B-N}$ BN. This corresponds to {\it symmetric} capping mentioned in the text. Also Se on the top of N on both sides of the film corresponds to {\it symmetric} case.}
\label{fig:Fig2}
\end{figure}

\begin{figure}[ht!]
\scalebox{0.53}{\includegraphics{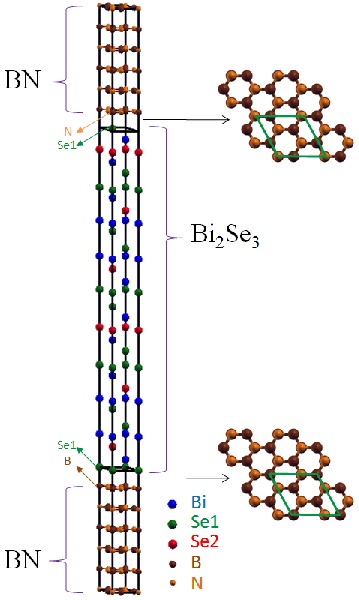}}
\caption{(Color online) Same as Fig. 2 but now the B on the top of Se on one side and N on the top of Se on the other side of the TI thin film. The colors corresonding to atoms are labled on the bottom right of supercell. The green, brown, and orange arrows indicate the relative position of Se atom with respect to B (bottom) and N (top). The top views of top and bottom interfacial atomic layers are shown on the top right and bottom right of the supercell, respectively. The 1$\times$1 Bi$_2$Se$_3$ cell (green rhombus) is matched with 3d$_{B-N}$$\times$3d$_{B-N}$ BN. This corresponds to {\it asymmetric} capping in the text.}
\label{fig:Fig3}
\end{figure}

\begin{figure}[ht!]
\scalebox{0.53}{\includegraphics{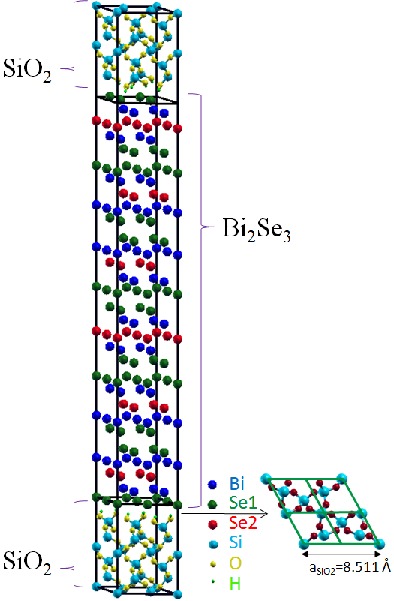}}
\caption{ (Color online) Schematic diagram of oxygen-terminated quartz with dangling bond states covered with hydrogen atoms (small green circles) interfaced on both sides of Bi$_2$Se$_3$. Si-terminated surface also possesses dangling bonds, but only oxygen-terminated surface without the hydrogen coverage distorts the surface state Dirac cone. The top view of bottom interfacial atomic layers is shown on the bottom right of the supercell structure. The 2$\times$2 Bi$_2$Se$_3$ cell (green rhombus) is matched with 1$\times$1 SiO$_2$ at the interface.}
\label{fig:Fig4}
\end{figure}

\section{Thin Films with Symmetric and Asymmetric Dielectric Cappings}

We chose crystalline SiO$_2$ in order to assess the effect of oxygen environment on the TI surface states. The choice of BN dielectrics is perhaps guided by the recent graphene transport experiments using crystalline dielectrics\cite{philip1, philip2}. We considered both {\it symmetric} and {\it asymmetric} cappings in terms of relative orientations of TI and dielectric surface atoms. {\it Asymmetry} in terms of different dielectrics on two opposite TI surfaces was not considered.  

The bulk structures of both dielectrics were studied with the computational parameters described in the previous section. For the hexagonal BN, the experimental lattice paramters are: {\it a}=0.2494 nm and {\it c}=0.666 nm with the distance between B and N d$_{B-N}$=0.144 nm\cite{CrystalBN}, and the band gap is 5.97 eV\cite{kanda}. We used the experimental values of lattice constants and atomic positions. With these experimental values, our calculated bulk band gap value is 5.5 eV close to the experimental value 5.97 eV. The hexagonal crystal structure of quartz contains fourfold coordinated oxygens, forming a layered structure with Si with the experimental lattice parameters ({\it a}=0.4914 nm and {\it c}=0.5408 nm\cite{quartz}). Our DFT calculations of optimized lattice paramters with semi-local potentials are found to be close to these experimental values, accurate to within 0.1$\%$. Therefore, we chose experimental lattice parameters for building our interface structures of TI and quartz. The bulk quartz SiO$_2$ has a direct band gap of 9 eV\cite{sio2gap}, and our DFT calculation of crystalline SiO$_2$ results in a band gap value of 9.4 eV. We note that our results obtained for quartz should be considered at best qualitative with regard to TIs capped with amorphous SiO2.

\subsection{Thin Films of Binary TI Bi$_2$Se$_3$ with Dielectrics}

In this section, we discuss the thin film of binary TI Bi$_2$Se$_3$.

\begin{figure}[ht!]
\scalebox{0.27}{\includegraphics[angle=-90]{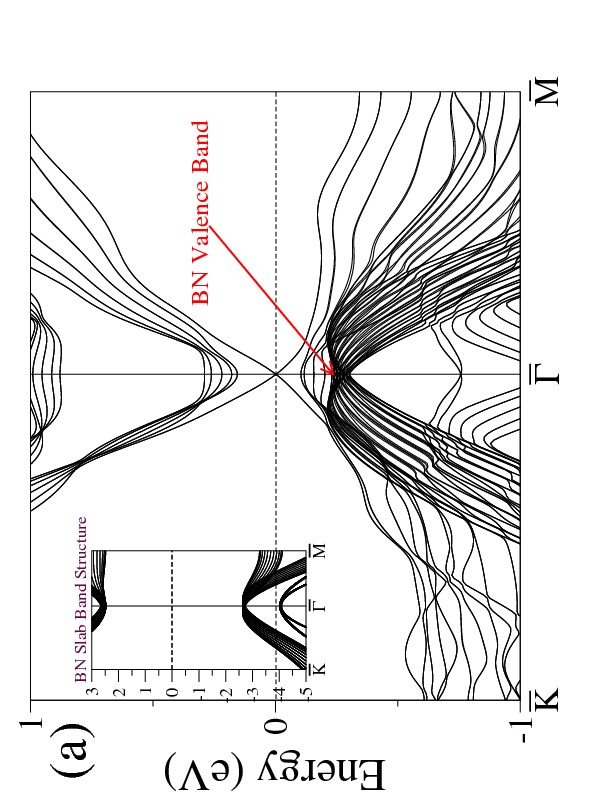}}
\scalebox{0.27}{\includegraphics[angle=-90]{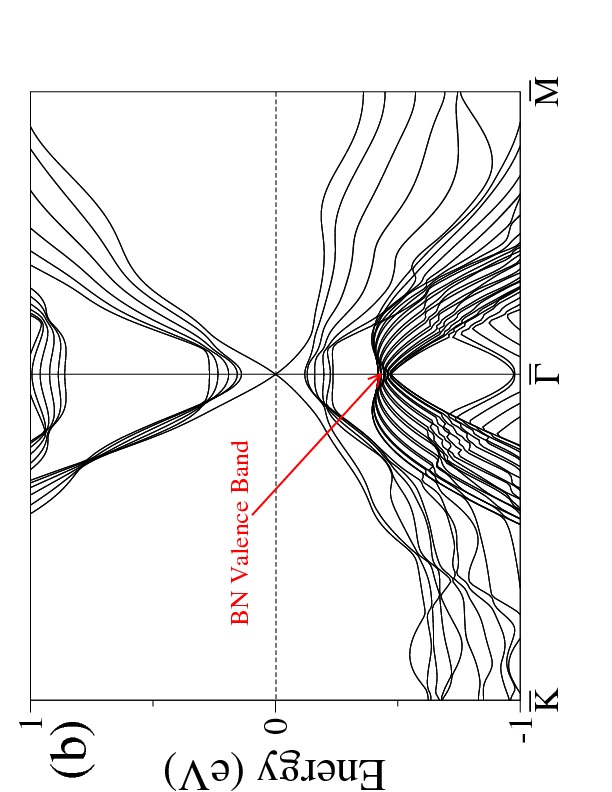}}
\scalebox{0.27}{\includegraphics[angle=-90]{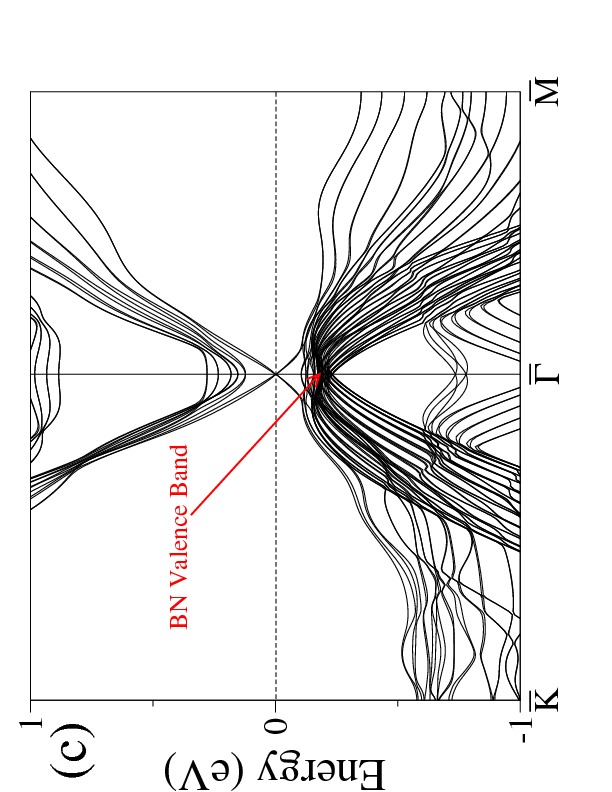}}
\caption{Band structures of Bi$_{2}$Se$_{3}$/BN supercell along high symmetry directions of the hexagonal BZ for {\it symmetric} capping case (a) B on the top of Se and (b) N on the top of Se on both sides of TI film without atomic relaxation. The Dirac cone and its degeneracy at the $\bar{\Gamma}$-point are not disturbed in the presence of crystalline BN thin film. (c) Same case of (a) with atomic relaxation. Atomic relaxation has little effect on the Dirac cone feature. The position of BN valence band maximum can be found by the band structure of BN slab without TI in the inset of (a) and the DOS plots in Fig. 6.}
\label{fig:Fig5}
\end{figure}

\begin{figure}[ht!]
\scalebox{0.27}{\includegraphics[angle=90]{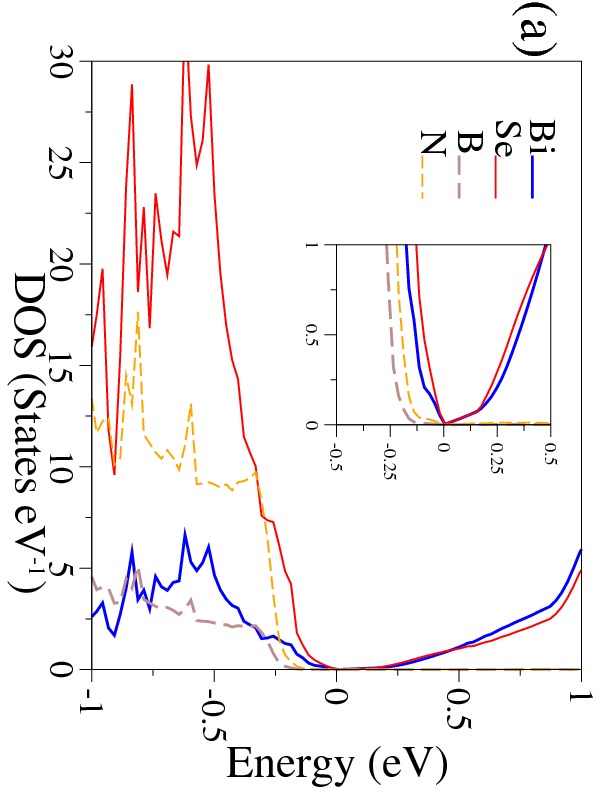}}
\scalebox{0.27}{\includegraphics[angle=90]{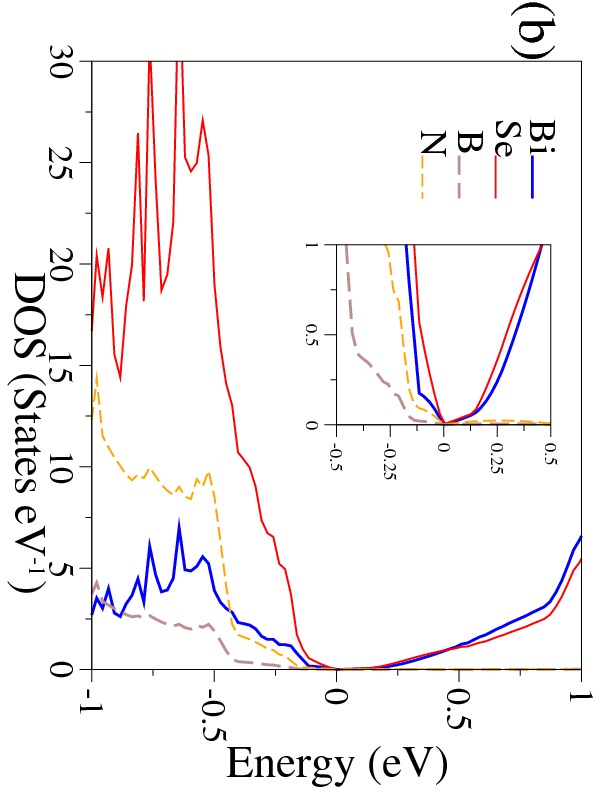}}
\caption{(Color online) The atom-projected DOS in the same energy range as the band structures in Fig. 5 for {\it symmetric} capping cases (a) B on the top of Se and (b) N on the top of Se on both sides of TI film without atomic relaxation. Se orbitals are not in {\it resonance} with either B or N orbitals near the Fermi level suggesting that the Dirac cone, formed from the Bi and Se orbitals, is intact.}
\label{fig:Fig6}
\end{figure}

\subsubsection{Construction of Supercell Structure}

The interface structures are built by putting the dielectric material on the surfaces of 6QLs of Bi$_2$Se$_3$ film stacked along the {\it z}-direction (since it has no band gap for the surface states). In constructing supercells, we maintained the TI lattice and strained the dielectrics to fit with the TI surface resulting in the compressive or tensile strain on the dielectrics to focus on their effects on the TI electronic structure. For the BN, our analysis suggests that to keep our computational burden in DFT-based calculations minimal, 3d$_{B-N}$$\times$3d$_{B-N}$ lattice structure can be matched in-plane with 1$\times$1 Bi$_2$Se$_3$ cell (Fig.2 and Fig.3) or any multiples of this combination which maintains 3:1 matching ratio resulting in the 4.19$\%$ compressive strain on BN (in Table I). Other combinations to allow less strain result in either difficulty in forming the periodic structure or larger supercells with orders of magnitude more atoms. Along the stacking direction, 6QLs of Bi$_2$Se$_3$ ($\sim$6nm) is put on six atomic layers of BN ($\sim$1.7nm). The choice of 6 BN layers is somewhat arbitrary and guided by the fact that the size of vacuum and BN layers should be thick enough to avoid interactions of periodically repeated Bi$_2$Se$_3$ surface layers. We also relaxed the interfacial atoms to assess its effects. For the SiO$_2$ dielectric, the supercell structure consists of two unit cells of SiO$_2$ sandwiching 6QLs Bi$_2$Se$_3$. The size of SiO$_2$ and Bi$_2$Se$_3$ cell along {\it x}-{\it y} direction chosen is, respectively, 1$\times$1 and 2$\times$2 (Fig. 4) to minimize the computational cost. This produces about 2.75$\%$ compressive strain on both sides of SiO$_2$ as in Table I. Consideration of larger sizes can lead to the lower strain, but the total number of atoms in the cell increases at least an order of magnitude (250 versus 2500). 

\begin{figure}[ht!]
\scalebox{0.27}{\includegraphics[angle=-90]{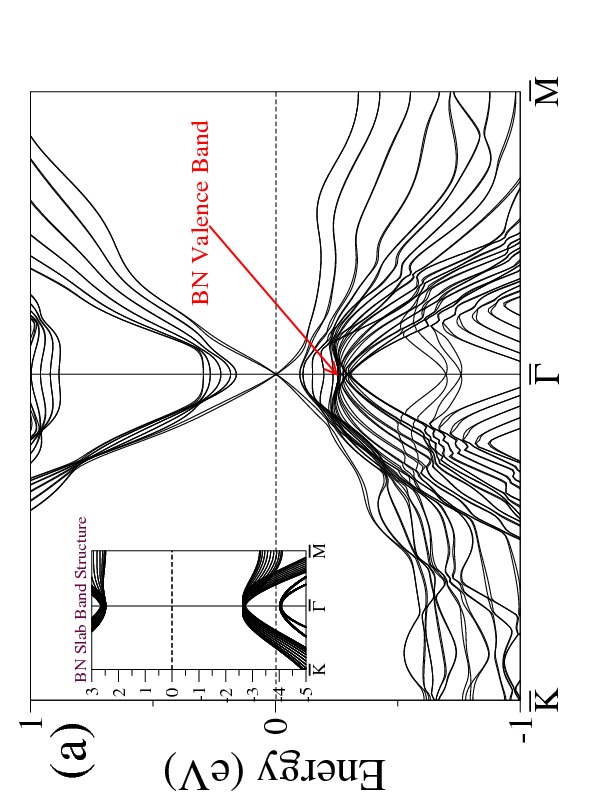}}
\scalebox{0.27}{\includegraphics[angle=-90]{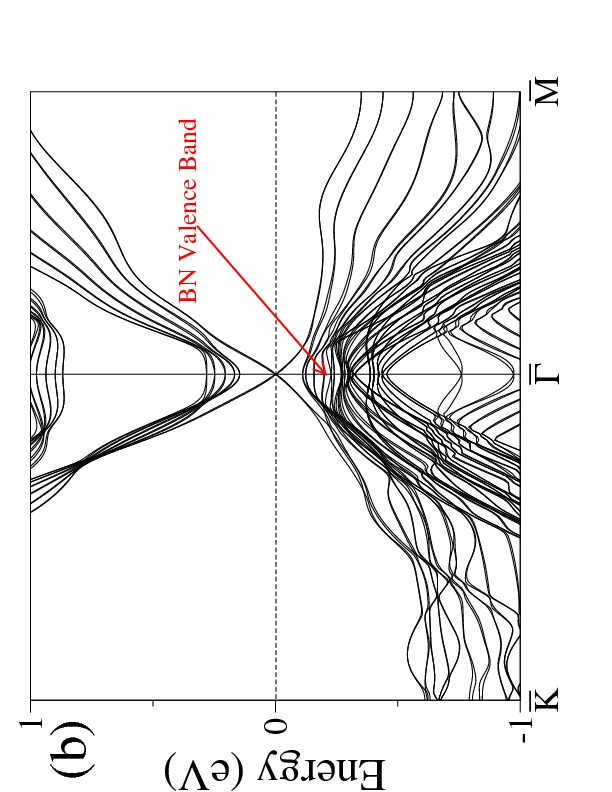}}
\caption{Band structures of Bi$_{2}$Se$_{3}$/BN supercell along high symmetry directions of the hexagonal BZ for {\it asymmetric} capping cases (a) B on the top of Se on one side and vacuum on the other side of TI film and (b) B on the top of Se atom on one side and N on the top of Se atom on the other side without atomic relaxation. The Dirac cone and its degeneracy at the $\bar{\Gamma}$-point are not disturbed in the presence of crystalline BN thin film. The position of BN valence band maximum can be found by the band structure of BN slab without TI in the inset of (a) and the DOS plots in Fig. 8.}
\label{fig:Fig7}
\end{figure}

\begin{figure}[ht!]
\scalebox{0.27}{\includegraphics[angle=90]{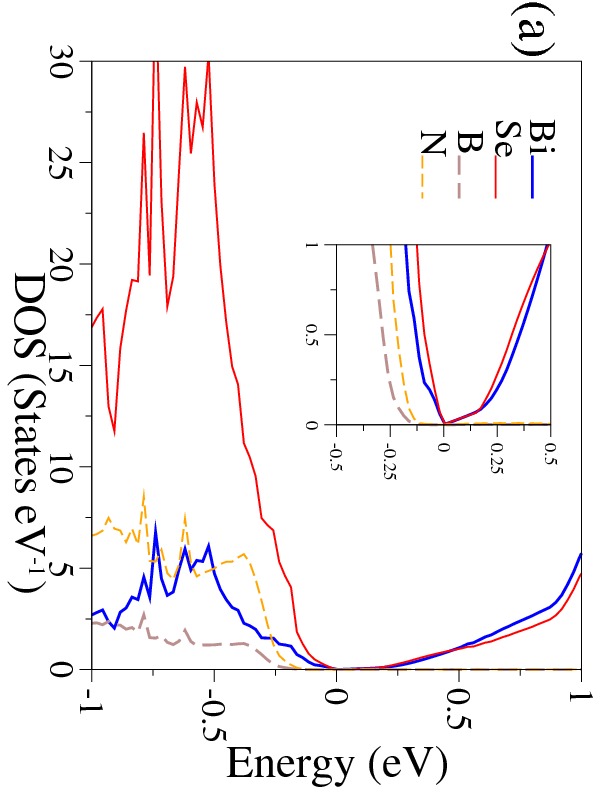}}
\scalebox{0.27}{\includegraphics[angle=90]{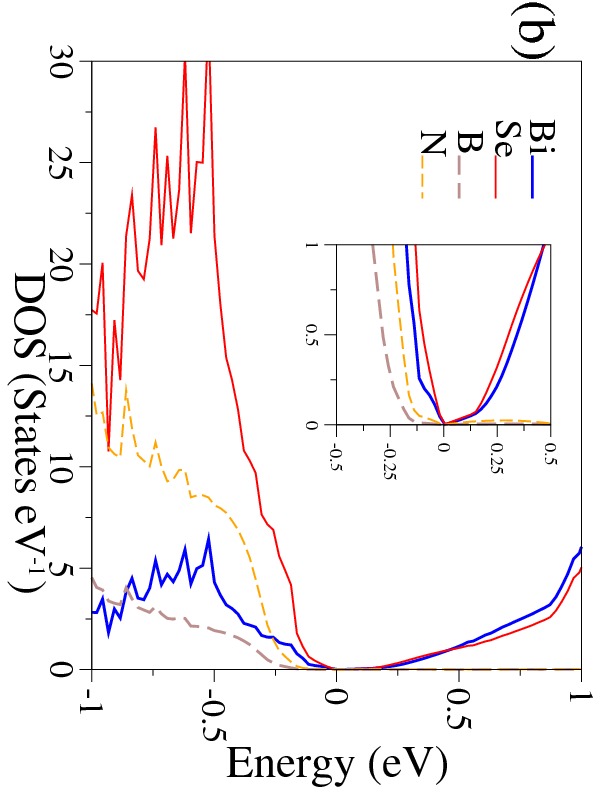}}
\caption{(Color online) The atom-projected DOS in the same energy range as the band structures in Fig. 7 for {\it asymmetric} capping cases (a) B on the top of Se on one side and vacuum on the other side of TI film and (b) B on the top of Se on one side and N on the top of Se on the other side of TI film without relaxation. Se orbitals are not in {\it resonance} with either B or N orbitals near the Fermi level suggesting that Dirac cone, formed from the Bi and Se orbitals, is intact.}
\label{fig:Fig8}
\end{figure}

We considered four configurations of Se positions on the TI surfaces with respect to boron and nitrogen positions on the BN layer: Se on the top of B, on the top of N, on the hexagonal hole and on the bond between B and N atoms. Two surface terminations, Si and oxygen, of quartz were considered. Our calculations suggests that all BN configurations are energetically close and provide quite similar band structures. Therefore, we had a choice in the selection of a particular configuration for further studies. We chose the case of B or N on the top of Se on both TI surfaces. We refer the structure of B or N on the top of Se on both sides of TI film as {\it symmetric} capping (Fig. 2). The structures of B on the top of Se on one side and N on the top of Se on the other side of TI film, or only B or N on one side of TI film and vacuum on the other side, are referred as {\it asymmetric} capping (Fig. 3). For quartz, both TI surfaces capped with either oxygen-terminated or Si-terminated quartz were considered. Because quartz is a fourfold coordinated structure, both of the surface terminations of quartz possess dangling bonds. Therefore, for quartz we also considered the dangling bond saturation with hydrogen, that is, hydrogen passivation of the quartz surfaces.

To set the optimal distance between the BN and Bi$_2$Se$_3$ layers at the interface, we performed the total energy calculations at the chosen set of interfacial distances. Our studies suggest an optimal distance of 0.3 nm for the case of the Se layer in Bi$_2$Se$_3$ on the top of either B or N in BN. For Bi$_2$Se$_3$ on quartz, the interfacial distances for Si and oxygen quartz terminations with and without hydrogen passivation need to be considered. Our studies suggest an optimal distance of 0.3 nm for Si-terminated surface, regardless of whether dangling bonds are saturated or not. For the oxygen-terminated quartz with hydrogen passivation, we found the optimal distance to be 0.25 nm. We considered these optimal interfacial distances in our further studies.

\begin{figure}[ht!]
\scalebox{0.27}{\includegraphics[angle=-90]{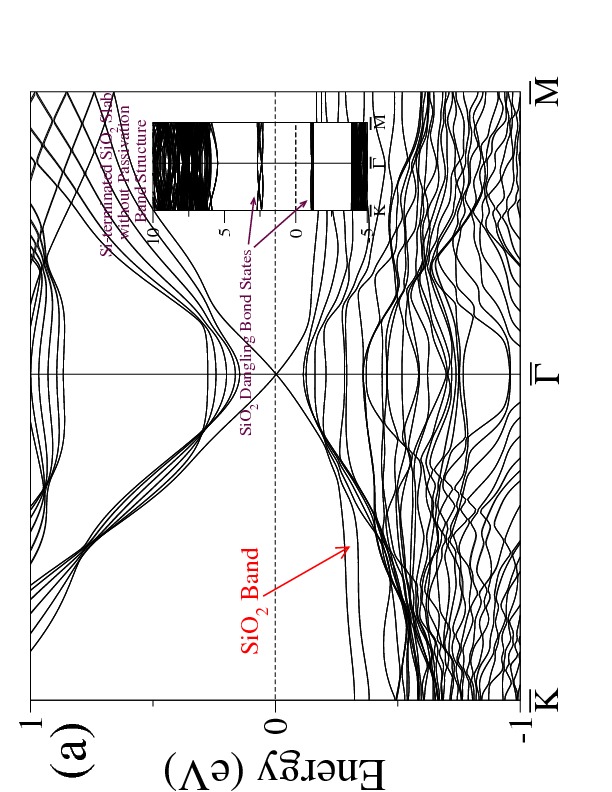}}
\scalebox{0.27}{\includegraphics[angle=-90]{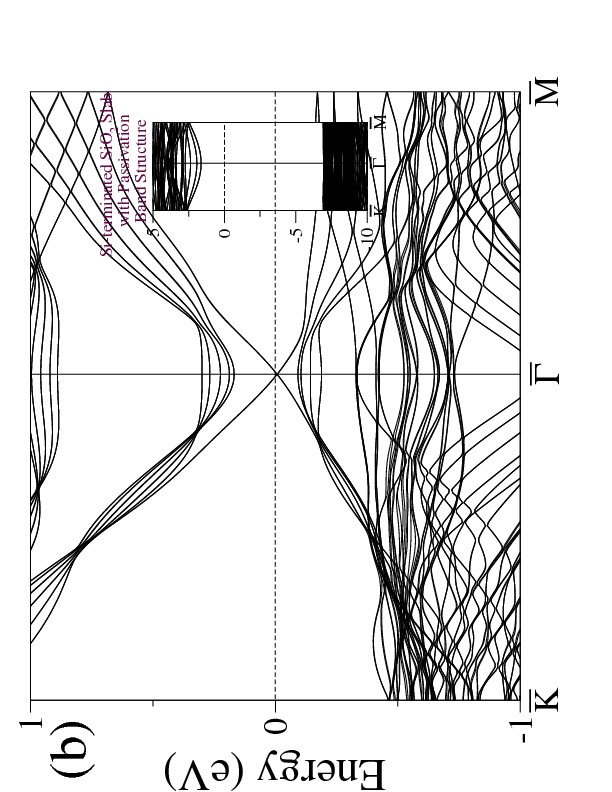}}
\caption{ Band structure of Bi$_{2}$Se$_{3}$ with Si-terminated quartz supercell along high symmetry directions in the hexagonal BZ (a) without hydrogen coverage and (b) with hydrogen coverage. The Dirac cone of TI is preserved in both cases. Insets of (a) and (b) show the band structures of only Si-terminated SiO$_2$ slab without and with passivation, respectively. In the inset of (b), a clear band gap is seen, while additional bands by Si dangling bond lie inside the gap in the inset of (a).}
\label{fig:Fig9}
\end{figure}

\begin{figure}[ht!]
\scalebox{0.27}{\includegraphics[angle=90]{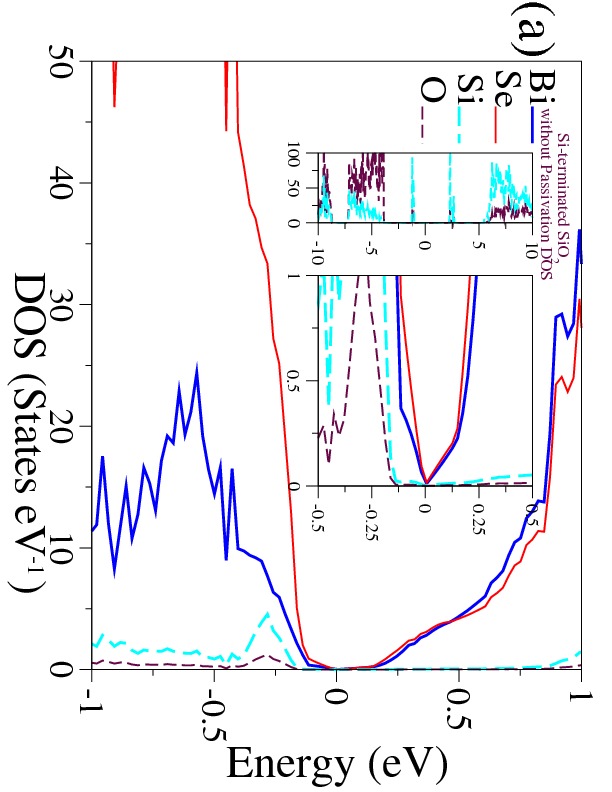}}
\scalebox{0.27}{\includegraphics[angle=90]{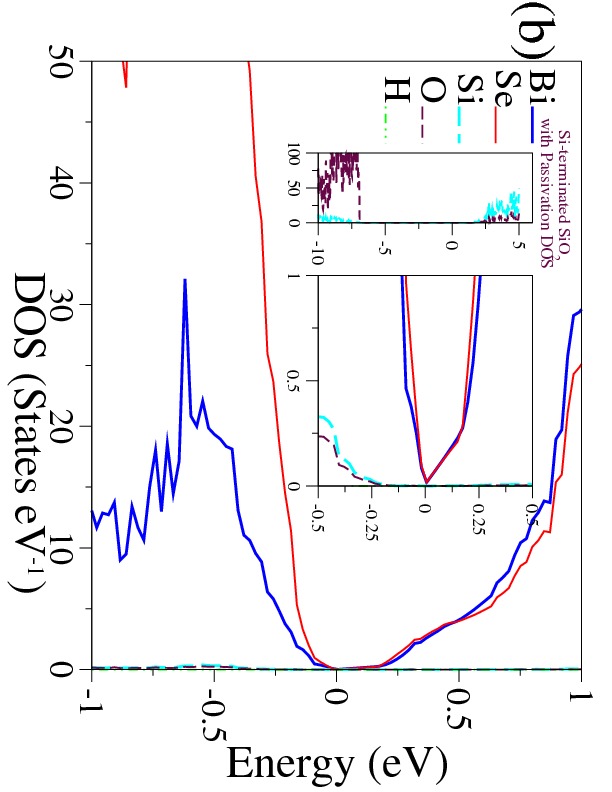}}
\caption{ (Color online) The atom-projected DOS in the same energy range as the band structures in Fig. 9 for the Si-terminated quartz cases (a) without hydrogen passivation and (b) with hydrogen passivation. The orbitals of Si and Se do not overlap near the Fermi level. Left insets of (a) and (b) show the atom-projected DOS for only Si-terminated SiO$_2$ slab without and with passivation, respectively. States from Si orbitals are clearly shown in the left inset of (a).}
\label{fig:Fig10}
\end{figure}

We considered the atomic relaxation in the interfacial region in order to check its effect on the electronic structure of TI surface states. Two cases were considered: first is the case of Se on the top of B atom on both sides of TI with the initial optimal separation of 0.3 nm. We let atoms close to the interface (Bi/Se atoms in the top and bottom QLs and B/N atoms in the two layers next to TI surfaces) move in the {\it z}-direction. The second choice is that the oxygen-terminated quartz with the saturation of dangling bonds is put on both TI surfaces with optimal separation of 0.25 nm. Bi/Se atoms in the top and bottom QLs as well as Si/O atoms next to the TI surfaces are allowed to move in the {\it z}-direction, while hydrogen atoms used to saturate dangling bonds are relaxed in all directions.

\begin{table}[b!]
\caption{Hexagonal unit cell lattice constant {\it a} of binary TI and strains on dielectrics by the lattice mismatch at the interface}
\begin{tabular}{ c  c  c  c }
\hline \hline
  TI & {\it a} (nm) & Strain on SiO$_{2}$ & Strain on BN \\
\hline
  Bi$_{2}$Se$_{3}$  & 0.41388 & 2.75$\%$ compressive & 4.19$\%$ compressive\\
\hline \hline
\end{tabular}
\end{table}

\begin{figure}[ht!]
\scalebox{0.27}{\includegraphics[angle=-90]{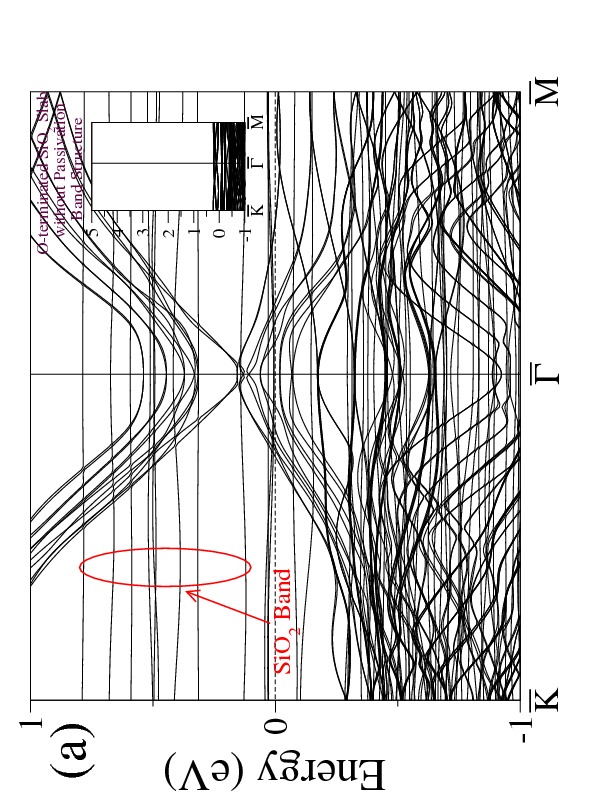}}
\scalebox{0.27}{\includegraphics[angle=90]{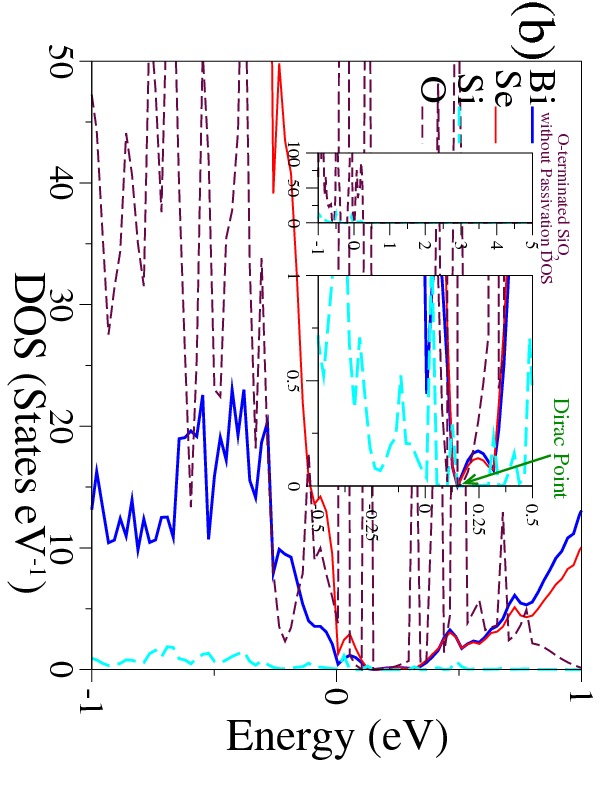}}
\caption{ (Color online) (a) Band structure of Bi$_{2}$Se$_{3}$ with oxygen-terminated quartz supercell without atomic relaxation along high symmetry directions in the hexagonal BZ suggesting that Dirac cone is significantly affected and (b) corresponding atom-projected DOS. The strong hybridization of Se and oxygen orbitals near the Fermi level is observed. Band structure of only oxygen-terminated SiO$_2$ slab without passivation is in the inset of (a), and corresponding the atom-projected DOS plot is shown in the left inset of (b). The Fermi level is below valence band edge formed by oxygen orbitals. The Dirac point is marked with the arrow on the right inset of (b).}
\label{fig:Fig11}
\end{figure}

\subsubsection{Results and Discussion}

We first discuss the BN interface effects followed by the effects due to quartz surface terminations. Figs. 5(a) and (b), respectively, show the band structures of 6QLs Bi$_2$Se$_3$ capped on both sides by BN with either B on the top of Se atom or N on the top of Se atom ({\it symmetric} capping). The atomic relaxation is not considered here. The Dirac cone and its degeneracy are protected in both cases. This insensitivity of the Dirac cone to the presence of dielectric hints at negligible interactions between B or N orbitals with Se orbitals. We confirm this hypothesis by plotting atom-projected DOS in the same energy range as the band structures in Figs. 6(a) and 6(b). These plots show that both B and N orbitals are not in {\it resonance} with the Se orbitals of TI. As a result, the Dirac cone surface states are protected. From the band structure of only BN atomic layers without Bi$_2$Se$_3$ (inset in Fig. 5(a)) and the atom-projected DOS plots of Figs. 6(a) and 6(b), the BN valence band maximum is estimated to be about 0.25 eV and 0.45 eV below the Dirac point for B on the top of Se atom and N on the top of Se atom cases, respectively. The band structure including the effect of atomic relaxation for the structure of B on the top of Se on both sides is shown in Fig. 5(c). Comparison Fig. 5(c) with Fig. 5(a) indicates that the relaxation of interfacial atomic positions does not affect the essential characteristics of surface states near the Fermi surface. Fourfold degeneracy at the Dirac point and the dispersion relation still remain same, while valence bands of BN are a bit shifted close to the Dirac point. The bulk band gap size for each case (0.259 eV for B on the top of Se, 0.258 eV for N on the top of Se) does not change much from  that of only 6QLs Bi$_2$Se$_3$ (0.262 eV).

In case of {\it asymmetric} capping, B on the top of Se atom on one side and vacuum on the other side of the TI film shows that the Dirac point degeneracy is still unaffected in Fig. 7(a). However, there is some splitting of valence and conduction bands close to it. With B on the top of Se on one side and N on the top of Se on the other side, the Dirac cone is not influenced either as seen in Fig. 7(b). Again, the atom-projected DOS plots in Figs. 8(a) and 8(b) suggest that this insensitivity is due to non-overlap of B and N orbitals with the Se orbitals. The position of BN valence band maximum is about 0.25 eV below the Fermi level for both cases similarly in the {\it symmetric} capping both TI surfaces with B on the Se, since B is placed on Se on one side of TI in both {\it asymmetric} cappings. The buulk band gap sizes are 0.268 eV and 0.257 eV for 'B on the top of Se atom on one side and vacuum on the other side' and 'B on the top of Se on one side and N on the top of Se on the other side', respectively.

\begin{figure}[ht!]
\scalebox{0.27}{\includegraphics[angle=-90]{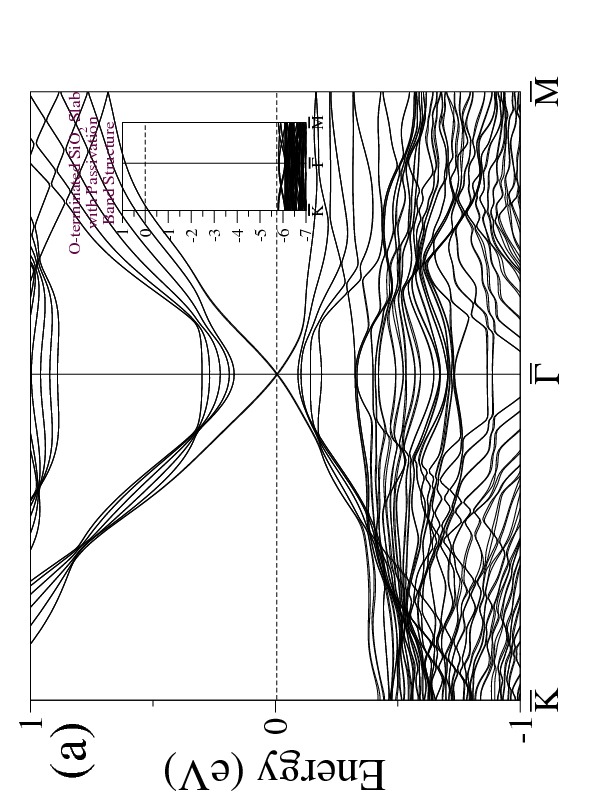}}
\scalebox{0.27}{\includegraphics[angle=90]{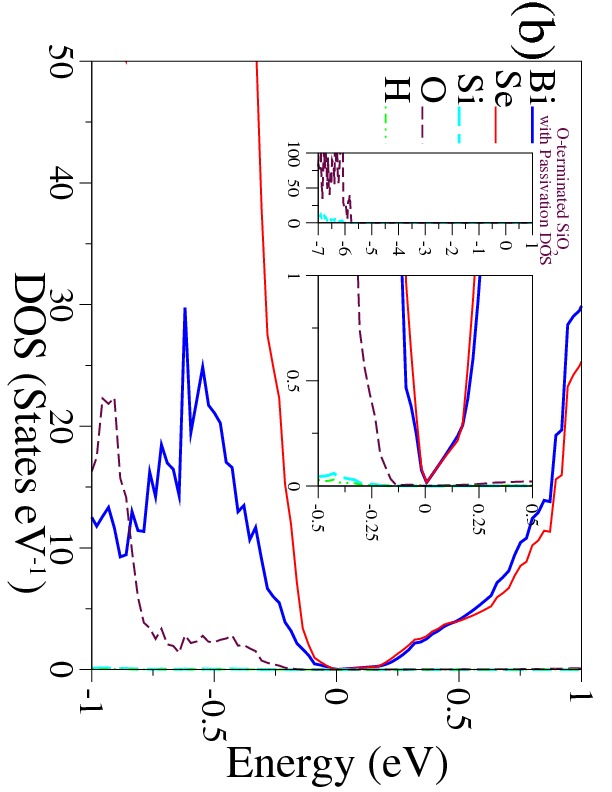}}
\scalebox{0.27}{\includegraphics[angle=-90]{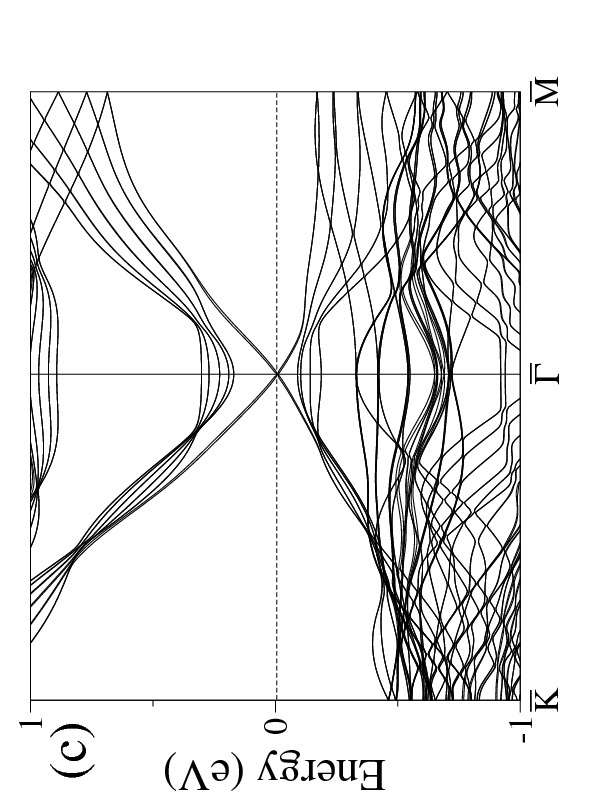}}
\caption{ (Color online) Same as Fig. 11 but now with hydrogen coverage of oxygen dangling orbitals (a) band structure without atomic relaxation and (b) corresponding atom-projected DOS. The Dirac cone of TI is recovered due to non interaction of Se and oxygen orbitals. (c) Same case with atomic relaxation. The Dirac cone feature is not affected by the atomic relaxation. Band structure of only oxygen-terminated SiO$_2$ slab with passivation is in the inset of (a), and corresponding the atomic-projected DOS plot is shown in the left inset of (b). The Fermi level is inside the gap.}
\label{fig:Fig12}
\end{figure}

For Si-terminated SiO$_2$, both without and with dangling bond passivations show the protected Dirac cone surface state (Figs. 9(a) and 9(b)), consistent with the DOS plots in Figs. 10(a) and 10(b). Si orbitals do not mix with Se orbitals in the energy range close to the Dirac point. The band structure of only Si-terminated quartz slab without the passivation (inset of Fig. 9(a)) and its DOS plot (left inset of Fig. 10(a)) show the Si dangling bond states within the bulk quartz band gap. These dangling bond states are also observed around the energy level of -0.26 meV in the band structure of Si-terminated quartz with Bi$_2$Se$_3$ (Fig. 9(a)), which is confirmed by the peak value of DOS from Si orbitals at -0.26 eV in Fig. 10(a). By the hydrogen passivation, dangling bond states are removed in the band gap of only Si-terminated SiO$_2$ Slab as seen in the inset of Fig. 9(b) and the left inset of Fig. 10(b). As a result, we cannot find the SiO$_2$ states in the energy range of -1$\sim$1 eV in Figs. 9(b) and 10(b). 

Fig. 11(a) is the band structure of the TI surface in the presence of oxygen-terminated quartz without passivation. The DOS plot (Fig. 11(b)) shows that oxygen orbitals lie close to the Se orbitals and dominate the region around Fermi level. By the effect of oxygen orbitals, the Dirac cone feature is significantly distorted. However, the Dirac point still remains intact, which can be confirmed by the linear increase of Bi and Se DOS around 0.15 eV in the right inset of Fig. 11(b). It exists close to the Fermi level and is buried inside SiO$_{2}$ valence band continuum as shown in Figs. 11(a) and 11(b). In the band structure of only oxygen-terminated quartz without hydrogen passivation, the Fermi level is under the valence band edge in the inset of Fig. 11(a). When oxygen-terminated quartz is put on the surface of Bi$_2$Se$_3$, the Fermi levels of two materials align, thereby pushing the Dirac point under the quartz valence band. Recent experimental studies of environmental effects on TI surface suggest the charge doping due to presence of oxygen\cite{shen}, but argue on the insensitiveness of TI Dirac cone. With oxygen surface passivation with atomic hydorgen, we recover a clear Dirac cone with the degeneracy point at the Fermi level (Fig. 12(a)) which is consistent with the DOS plots in Fig. 12(b). In the only oxygen-terminated quartz slab with passivation, we can observe that the Fermi level resides inside the band gap from the band structure and DOS plot shown in the inset of Fig. 12(a) and the left inset of Fig. 12(b), respectively. Therefore, the Fermi level matching between quartz and Bi$_2$Se$_3$ occurs without pulling down the Dirac cone surface states into the quartz valence band. Atomic relaxations of interfacial atoms has no effect on the Dirac cone as well as its degeneracy (Fig. 12(c)). For all SiO$_2$ cappings, except oxygen-terminated quartz without passivation, the Dirac cones reside inside the bulk band gap of about 0.26 eV.
     
\subsection{Thin Films of Ternary TIs Bi$_2$Se$_2$Te and Bi$_2$Te$_2$Se}

In this section, we address the dielectric effects on the electronic structure of ternary TIs of Bi and Chalcogen. 

\subsubsection{Construction of Supercell Structure}

The construction of interface structures for ternary TI Bi$_2$Se$_2$Te (Bi$_2$Te$_2$Se) is similar with that for Bi$_2$Se$_3$ except the thickness of TI films. Since 4QLs is predicted as a minimum thickness to preserve the Dirac cone band structure for ternary TIs\cite{jiwon}, 4QLs thick TI films are used in the ternary TI calculations instead of 6QLs in the interface structure of binary TI Bi$_2$Se$_3$ with dielectrics.

\begin{table} [b!]
\caption{Hexagonal unit cell lattice constant {\it a} of ternary TIs and strains on dielectrics by the lattice mismatch at the interface}
\begin{tabular}{ c  c  c  c }
\hline \hline
  TI & {\it a} (nm) & Strain on SiO$_{2}$ & Strain on BN \\
\hline
  Bi$_{2}$Se$_{2}$Te  & 0.422 & 0.83$\%$ compressive & 0.23$\%$ compressive\\
\hline
  Bi$_{2}$Te$_{2}$Se  & 0.428 & 0.57$\%$ tensile & 0.92$\%$ compressive\\
\hline \hline
\end{tabular}
\end{table}

\begin{figure}[ht!]
\scalebox{0.27}{\includegraphics[angle=-90]{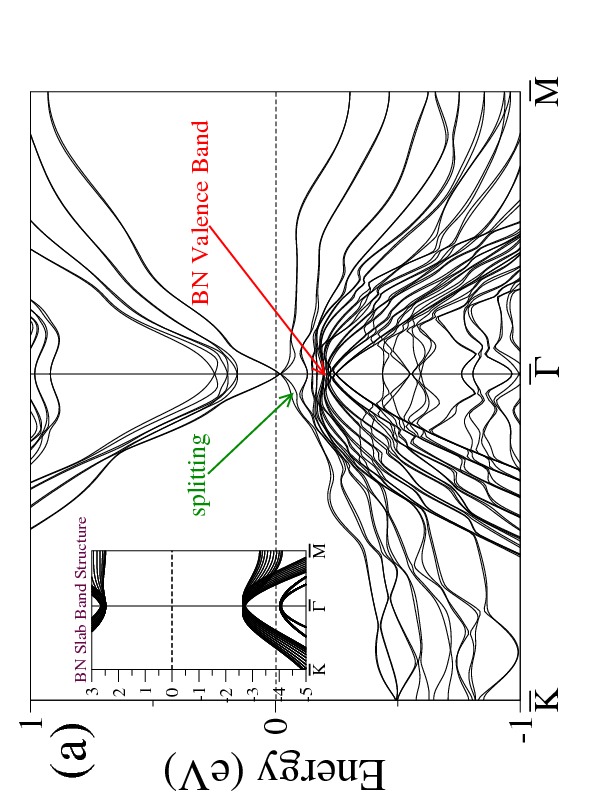}}
\scalebox{0.27}{\includegraphics[angle=-90]{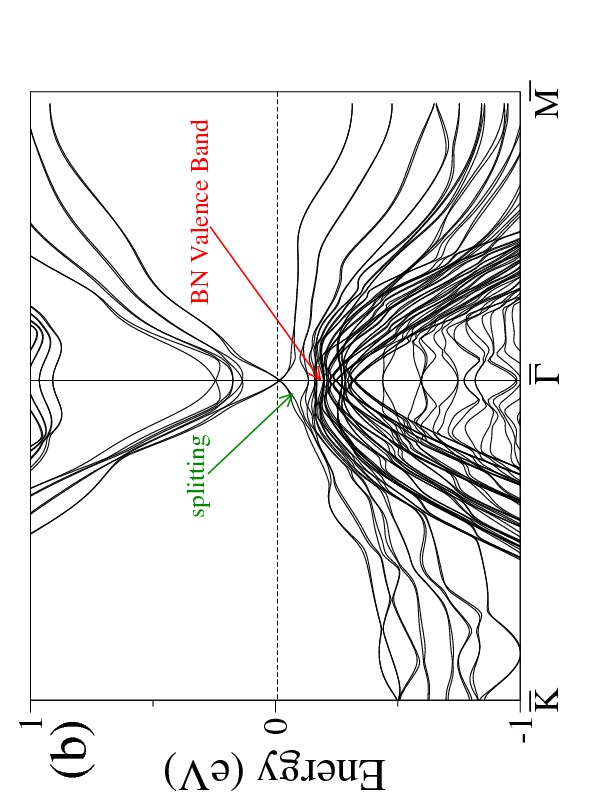}}
\scalebox{0.27}{\includegraphics[angle=-90]{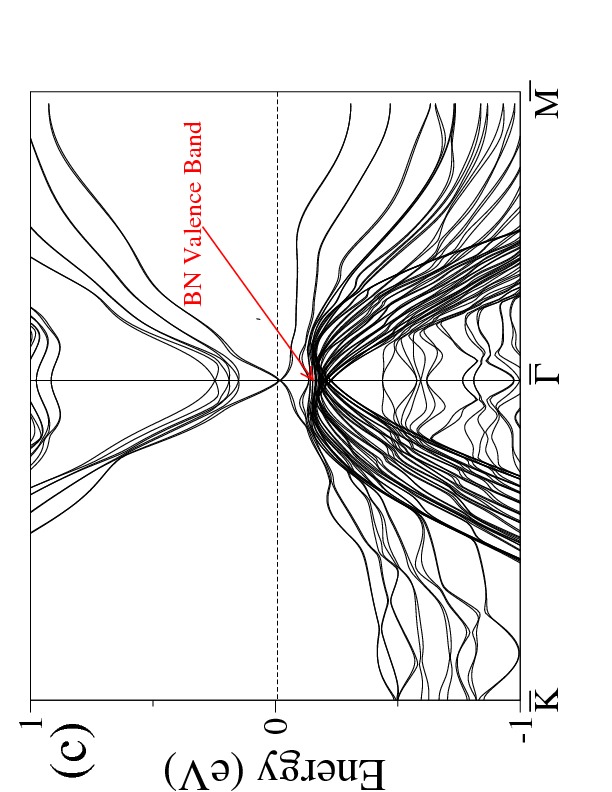}}
\caption{Band structure of Bi$_{2}$Se$_{2}$Te/BN supercell along high symmetry directions of the hexagonal BZ for {\it asymmetric} capping cases (a) B on the top of Se on one side and vacuum on the other side of TI film and (b) B on the top of Se atom on one side and N on the top of Se atom on the other side and for {\it symmetric} capping case (c) B on the top of Se on both sides of TI film. The atomic relaxation is not considered. The Dirac cone and its degeneracy at the $\bar{\Gamma}$-point are not disturbed in the presence of crystalline BN thin film. The position of BN valence band edge can be estimated by the band structure of BN slab without TI in the inset of (a).}
\label{fig:Fig13}
\end{figure}

For the BN dielectric capping, the BN of 3d$_{B-N}$$\times$3d$_{B-N}$ lattice structure is matched in-plane with the 1$\times$1 Bi$_2$Se$_2$Te (Bi$_2$Te$_2$Se) cell. Since we keep the lattice constant of Bi$_2$Se$_2$Te (Bi$_2$Te$_2$Se) and fit BN into it, BN is under about 0.23$\%$ compressive strain (0.92$\%$ compressive strain) in Table II. Six layers of BN ($\sim$1.7nm) is put on each side of 4QLs Bi$_2$Se$_2$Te (Bi$_2$Te$_2$Se). For the SiO$_2$ dielectric, the 1$\times$1 SiO$_2$ cell is fit with the 2$\times$2 Bi$_2$Se$_2$Te (Bi$_2$Te$_2$Se) cell in-plain, which results in 0.83$\%$ compressive strain (0.57$\%$ tensile strain) on SiO$_2$ as summarized in Table II. Two unit cells of SiO$_2$ is stacked on both sides of 4QLs Bi$_2$Se$_2$Te (Bi$_2$Te$_2$Se) in the {\it z}-direction.

\begin{figure}[ht!]
\scalebox{0.27}{\includegraphics[angle=-90]{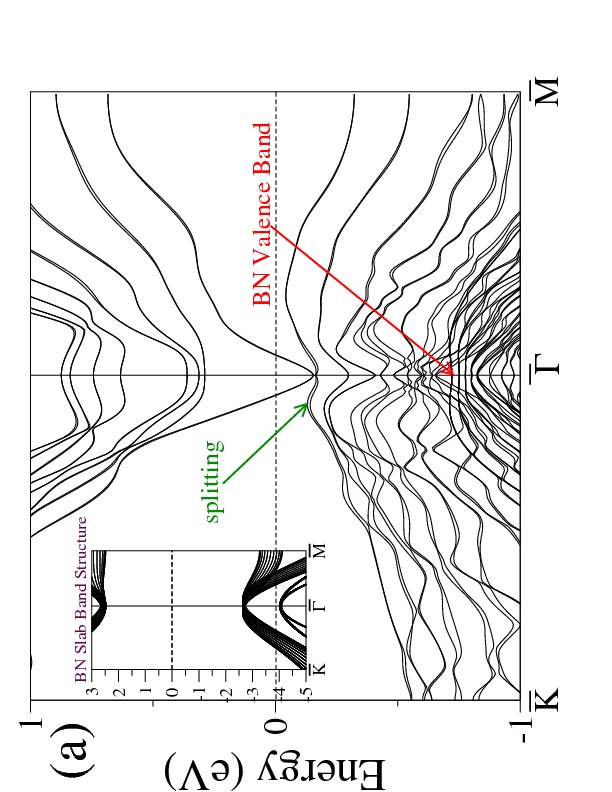}}
\scalebox{0.27}{\includegraphics[angle=-90]{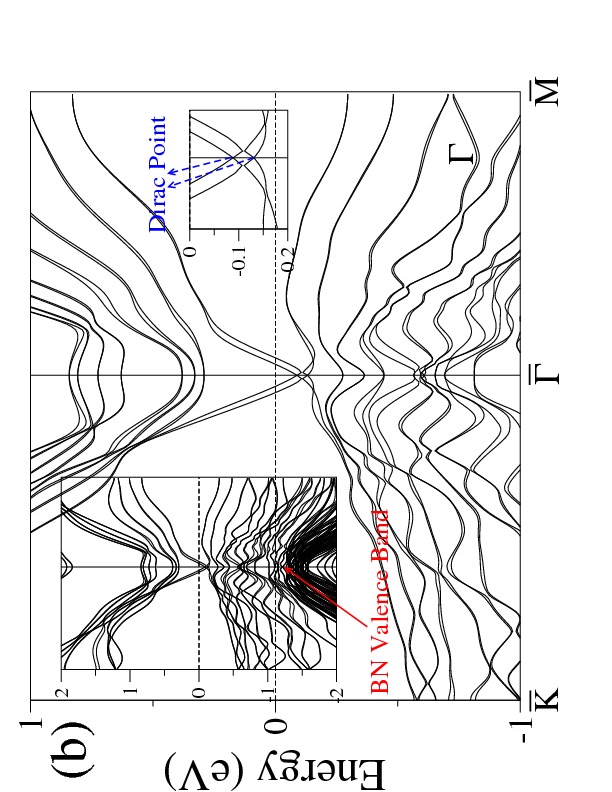}}
\scalebox{0.27}{\includegraphics[angle=-90]{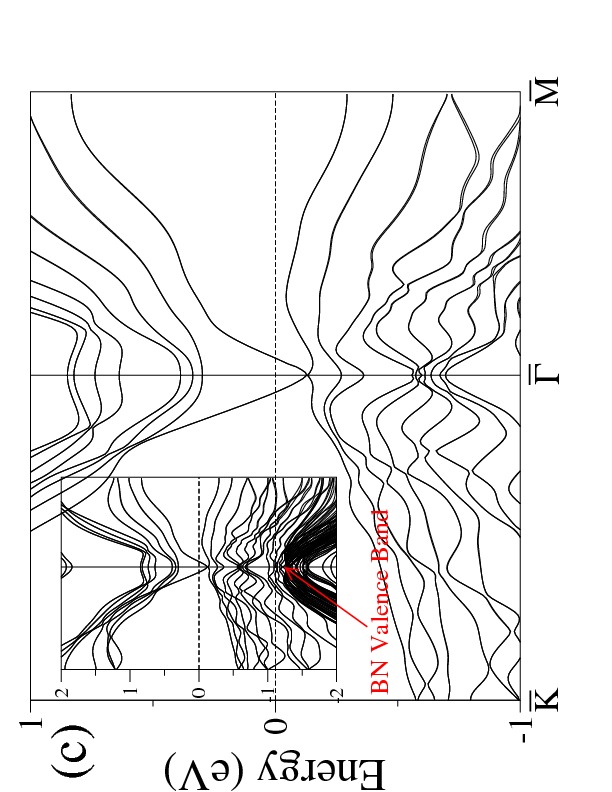}}
\caption{Band structure of Bi$_{2}$Te$_{2}$Se/BN supercell along high symmetry directions of the hexagonal BZ for {\it asymmetric} capping cases (a) B on the top of Te on one side and vacuum on the other side of TI film and (b) B on the top of Te atom on one side and N on the top of Te atom on the other side and for {\it symmetric} capping case (c) B on the top of Te on both sides of TI film. The atomic relaxation is not considered. The Dirac cone at the $\bar{\Gamma}$-point is not disturbed in the presence of crystalline BN thin film. The fourfold degeneracy is lifted to two twofold degeneracies at the $\bar{\Gamma}$-point in the inset of (b). The position of BN valence band maximum can be found by the band structure of BN slab without TI in the inset of (a).}
\label{fig:Fig14}
\end{figure}

\begin{figure}[ht!]
\scalebox{0.27}{\includegraphics[angle=-90]{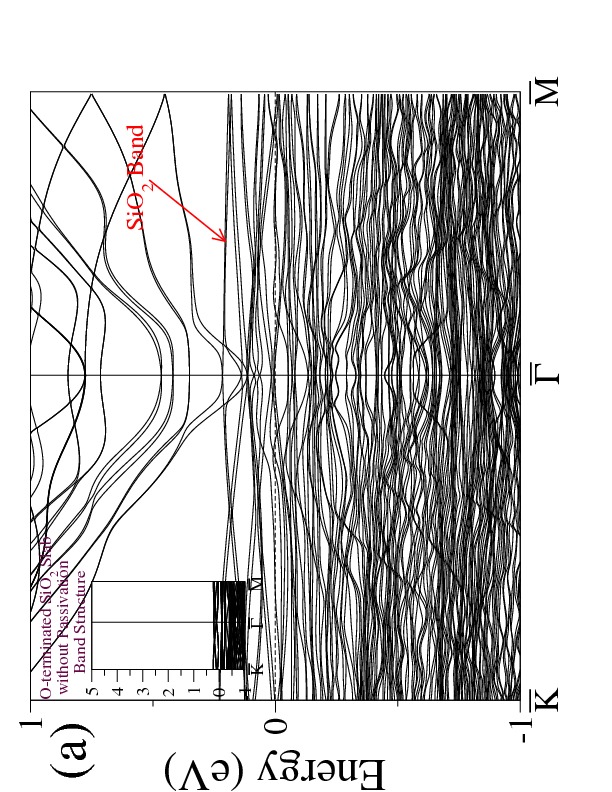}}
\scalebox{0.27}{\includegraphics[angle=90]{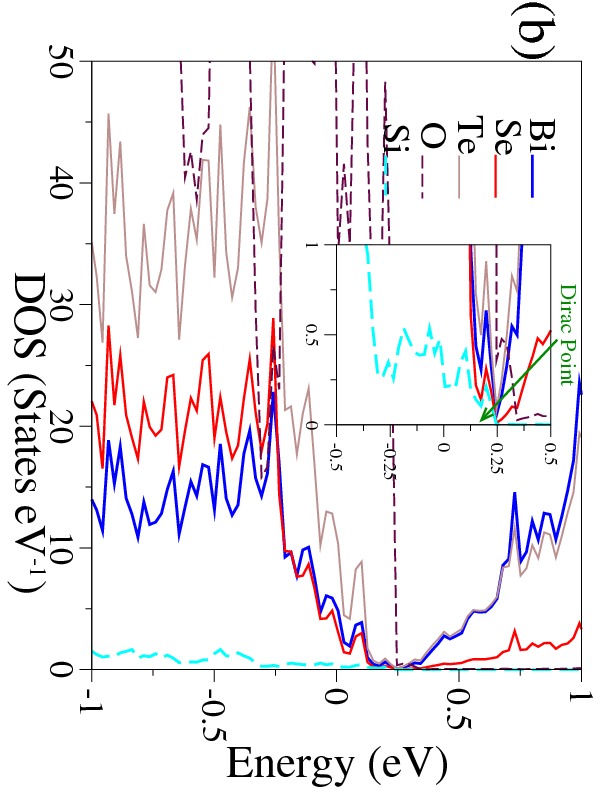}}
\caption{ (Color online) (a) Band structure of Bi$_{2}$Te$_{2}$Se with oxygen-terminated quartz supercell without atomic relaxation along high symmetry directions in the hexagnoal BZ suggesting that Dirac cone is affected and (b) corresponding atom-projected DOS. The Dirac point of TI is buried into quartz valence bands. Band structure of only oxygen-terminated SiO$_2$ slab without passivation is in the inset of (a). The Fermi level is below valence band edge formed by oxygen orbitals. The Dirac point is marked with the arrow on the inset of (b).}
\label{fig:Fig15}
\end{figure}

Four configurations of Se (Te) positions on the TI surfaces with respect to B and N positions on the BN layer were investigated and found to be energetically quite similar, same as our studies on the binary TI with BN. The optimal distance between BN and Bi$_2$Se$_2$Te (Bi$_2$Te$_2$Se), minimizing the total energy, is found to be 0.32 nm (0.3 nm). We considered {\it asymmetric} capping, B on the top of Se (Te) on one side and vacuum on the other side or B on the top of Se (Te) on one side and N on the top of Se (Te) on the other side, as well as {\it symmetric} capping, B on the top of Se (Te) on both sides. For the SiO$_2$ capping, we focused on the oxygen-terminated SiO$_2$ on both sides of TI without saturating oxygen dangling bond states as a critical case because there are reports, as discussed in the previous section, that oxygen dangling bond states may play a crucial role in modifying TI surface. Only Bi$_2$Te$_2$Se with oxygen-terminated quartz was studied because we know that Se orbitals lie close in energy with the oxygen orbitals from the previous study of Bi$_2$Se$_3$ and the ternary TI Bi$_2$Se$_2$Te is structurally similar to the binary TI Bi$_2$Se$_3$ regarding the atom species Se of top and bottom surfaces. Atomic relaxation was not considered for ternary TIs because the rearrangement of atomic positions was found to have no substantial effect in the binary TI with dielectrics.

\subsubsection{Results and Discussion}

Computed band structures of Bi$_2$Se$_2$Te/BN supercell are seen in Fig. 13. Both {\it asymmetric} capping, in Fig. 13(a) (B on the top of Se on one side and vacuum on the other side) and Fig. 13(b) (B on the top of Se on one side and N on the top of Se on the other side), and {\it symmetric} capping in Fig. 13(c) (B on the top of Se on both sides) indicate that the Dirac point and the surface state dispersion are preserved. However, small splitting between top and bottom surface bands occurs in the {\it asymmetric} case, which are induced by different environments on opposite surfaces, while no such splitting exists in the {\it symmetric} structure. The valence band maximum is about 0.25 eV below for all three cases as we can estimate from the band structure of only BN slab in the inset of Fig. 13(a) and atom-projected DOS plots (figures not shown). The bulk band gap sizes (around 0.251 eV) in all BN cappings for Bi$_2$Se$_2$Te increases a little bit as compared with the bulk gap (0.204 eV) of only 4QLs Bi$_2$Se$_2$Te.

Fig. 14 shows band structures of Bi$_2$Te$_2$Se/BN supercell. In Fig. 14(a), the {\it asymmetric} capping of B on the top of Te on one side and vacuum on the other side, the fourfold degeneracy at the Dirac point is maintained, but two opposite surface bands seem to split. Another {\it asymmetric} capping that Te is on the top of B on one side and N on the other side leads to the split of fourfoud degeneracy into two twofold degeneracies at the $\bar{\Gamma}$-point. The twofold degeneracy on each surface band still remains due to Kramer's theorem which requires that the twofold degeneracy at the time-reversal invariant momenta points is protected in the absence of time-reversal symmetry breaking perturbations. On the other hands, in the {\it symmetric} structure of B on the top of Te on both sides, surface bands on both sides are perfectly aligned as seen in Fig. 14(c). In BN cappings of Bi$_2$Te$_2$Se, the bulk band gap sizes (around 0.33 eV) are a bit larger than 0.325 eV the bulk band gap of 4QLs Bi$_2$Te$_2$Se without BN.

For the capping Bi$_2$Te$_2$Se with the oxygen-terminated quartz without dangling bond passivation, the Dirac cone surface states are highly influenced by the oxygen dangling bond states in Fig. 15(a). From the DOS plot in Fig. 15(b), Bi, Te and Se orbitals are overlapped with oxygen orbitals near the Dirac point. The Dirac point indicated in the inset in Fig. 15(b) is within the valence bands of quartz. The DOS values for Bi, Se and Te at the Dirac point are not zero as shown in the inset because the Dirac point is already buried under the bulk valence band maximum of Bi$_2$Te$_2$Se\cite{jiwon}. Similarly with Bi$_2$Se$_3$, oxygen dangling states modify Dirac cone surface states without breaking the Dirac point. 



\section {Summary and Conclusions}
We use a density functional based electronic structure method and atom-projected DOS to study the perturbations from dielectric cappings to the Dirac cone surface states of Bi-based binary and ternary TIs. Two crystalline dielectrics BN and quartz were considered with both {\it symmetric} and {\it asymmetric} cappings. Our study suggests that oxygen coverage substantially affects the Dirac cone, consistent with a recent experimental study. All other surface dielectric terminations have no significant effect on the TI surface states.  

\acknowledgments
The authors acknowledge financial support from the Nanoelectronics Research Initiative supported Southwest Academy of Nanoelectronics (NRI-SWAN) center. We thank Texas advanced computing center (TACC) for computational support (TG-DMR080016N).

\end{document}